\documentclass{PoSmod}
\usepackage[utf8]{inputenc}
\usepackage{amssymb}
\usepackage{slashed}
\usepackage{amsbsy}		
\usepackage{amsfonts}		
\usepackage{amsmath}		
\usepackage{url}		

\newcommand{\kd}{\kappa^{D}}
\newcommand{\rd}{\rho^{D}}
\newcommand{\ku}{\kappa^{U}}
\newcommand{\ru}{\rho^{U}}
\newcommand{\kl}{\kappa^{L}}
\newcommand{\rl}{\rho^{L}}

\newcommand{\sba}{s_{\beta-\alpha}}
\newcommand{\cba}{c_{\beta-\alpha}}

\title{2HDMC -- a two Higgs Doublet Model Calculator}

\ShortTitle{2HDMC -- a two Higgs Doublet Model Calculator}

\author{\speaker{Johan Rathsman}\thanks{On leave of absence from Uppsala University.}\\
        Department of Astronomy and Theoretical Physics,  Lund University\\ 
S{\"o}lvegatan 14A, SE-223 62 Lund, Sweden \\
        E-mail: \email{Johan.Rathsman@thep.lu.se}}
\ReportNo{DESY 11-054\\ LU-TP 11-16} 

\author{Oscar St{\aa}l\\
        Deutsches Elektronen-Synchrotron DESY\\ Notkestra{\ss}e 85, D-22607 Hamburg, Germany\\
       E-mail: \email{oscar.stal@desy.de}}

\abstract{We present the program 2HDMC and how it can be used to explore the physics of general CP-conserving two Higgs doublet models.}

\FullConference{Third International Workshop on Prospects for Charged Higgs Discovery at Colliders - CHARGED2010,\\
		September 27-30, 2010\\
		Uppsala Sweden}

\begin{document}

\section{Introduction}
Two Higgs doublets models (2HDM) is one of the simplest, non-trivial extensions of the standard model Higgs sector, which appears for example in the Minimal SuperSymmetric Model (MSSM). For a general introduction to 2HDMs and the Higgs sector of MSSM we refer to \cite{HHG,Djouadi:2005gj}. A general 2HDM can therefore serve as an effective theory when searching for physics beyond the standard model. In this contribution we describe the program 2HDMC~\cite{2HDMC} and how it can be used to explore the physics of general CP-conserving 2HDMs \footnote{Admittedly, given the format for these proceedings, this description will have to be quite limited and incomplete and we apologize for that. For a more complete description we refer to ~\cite{2HDMC} and references therein.}.

\section[2HDM]{General two Higgs doublet models}
 \subsection[2HDM potential]{General two Higgs doublet model potential}
 We consider the  potential for a general model with two complex $SU(2)_L$ doublets 
 with hypercharge Y=1: $\Phi_1$ ,$\Phi_2$ which is invariant under global $SU(2)$ 
transformations, $\Phi_a \to U_{ab}\Phi_b$, and that is gauge invariant and renormalizable
 \begin{equation}
  \begin{aligned}
    \mathcal{V} = & m_{11}^2\Phi_1^\dagger\Phi_1+m_{22}^2\Phi_2^\dagger\Phi_2
    -\left[m_{12}^2\Phi_1^\dagger\Phi_2+\mathrm{h.c.}\right]
        \\
+&\frac{1}{2}\lambda_1\left(\Phi_1^\dagger\Phi_1\right)^2
    +\frac{1}{2}\lambda_2\left(\Phi_2^\dagger\Phi_2\right)^2
    +\lambda_3\left(\Phi_1^\dagger\Phi_1\right)\left(\Phi_2^\dagger\Phi_2\right)
    +\lambda_4\left(\Phi_1^\dagger\Phi_2\right)\left(\Phi_2^\dagger\Phi_1\right)
    \\+&\left\{
    \frac{1}{2}\lambda_5\left(\Phi_1^\dagger\Phi_2\right)^2
    +\left[\lambda_6\left(\Phi_1^\dagger\Phi_1\right)
      +\lambda_7\left(\Phi_2^\dagger\Phi_2\right)
      \right]\left(\Phi_1^\dagger\Phi_2\right)
    +\mathrm{h.c.}\right\} \nonumber
  \end{aligned}
  \label{eq:pot_gen}
\end{equation}
Demanding that the   
   potential is real implies that {$m_{11}^2$, $m_{22}^2$, $\lambda_{1-4}$} are real  
whereas {$m_{12}^2$, $\lambda_{5-7}$} in general can be complex. However, if we demand that there is 
no explicit CP-violation then the latter parameters also have to be real.

 \subsection[EWSB]{Electroweak symmetry breaking}
As usual the electroweak symmetry is broken by non-zero vacuum expectation values (vev) of $\Phi_1$ and/or $\Phi_2$. After applying the minimization conditions of the potential the $m_{11}^2$ and $m_{22}^2$ parameters can be traded for the vev's of the two doublets: $v_1=v\cos\beta$ and $v_2=ve^{i\xi}\sin\beta$ with $v=(\sqrt{2}G_F)^{-1/2}\approx246$ GeV. Here $\xi$ is a possible phase that we put to zero by demanding that there is no  spontaneous CP-violation. The two doublets can then be written as

\begin{equation}
\Phi_1=\frac{1}{\sqrt{2}}\left(\begin{array}{c}
\displaystyle \sqrt{2}\left(G^+\cos\beta -H^+\sin\beta\right)  \\
\displaystyle v\cos\beta-h\sin\alpha+H\cos\alpha+\mathrm{i}\left( G^0\cos\beta-A\sin\beta \right)
\end{array}
\right) \nonumber
\end{equation}
\begin{equation}
\Phi_2=\frac{1}{\sqrt{2}}\left(\begin{array}{c}
\displaystyle \sqrt{2}\left(G^+\sin\beta +H^+\cos\beta\right)  \\
\displaystyle v\sin\beta+h\cos\alpha+H\sin\alpha+\mathrm{i}\left( G^0\sin\beta+A\cos\beta \right)
\end{array}
\right)  \nonumber
\end{equation}
where $G^+$ and $G^0$ are the Nambu-Goldstone bosons that give masses to the $W$ and $Z$ bosons respectively, $H^+$ is the charged Higgs boson, $A$ is a CP-odd and $h$, $H$ are CP-even neutral Higgs bosons (with $m_h \leq m_H$), and $\tan\beta$ defines a basis in $\Phi$ space. (Also note that there is a special basis, the Higgs basis, where only one of the fields develops a vev and therefore $\tan\beta=0$ or $\cot\beta=0$). With these conventions the couplings of Higgs and electroweak gauge bosons  are given by the invariant $s_{\beta-\alpha}\equiv \sin(\beta-\alpha)$ and  $c_{\beta-\alpha}\equiv \cos(\beta-\alpha)$. Finally we note that in addition to the above parameterisation of the potential it can also be useful to use the masses of the Higgs bosons as parameters. Thus in the 2HDMC program the user can choose between the so called general parameterisation,
   \{$m_{12}^2$, $\lambda_{1-7}$, $\tan\beta$\} or the physical one
    \{$m_{12}^2$, $m_h$, $m_H$, $m_A$, $m_{H^\pm}$, $s_{\beta-\alpha}$, $\lambda_{6-7}$, $\tan\beta$\} in addition to the Higgs basis which is the same as the general one but with $\tan\beta=0$ and $m_{H^\pm}$ replacing $m_{12}^2$. For backward compatibility the program can also handle the basis used in the Higgs Hunter's Guide~\cite{HHG}. It is important to note that the program uses conventions where $\tan\beta\geq0$ and $s_{\beta-\alpha}\geq0$.

 \subsection[Symmetries]{Possible additional symmetries}
In addition to the symmetries already discussed it is also possible to impose other interesting symmetries on the potential.
First we note that one can demand an additional $U(1)_{PQ}$ (Peccei-Quinn) symmetry  \cite{Peccei:1977hh} by setting  $m_{12}^2=0$ and $\lambda_{5-7} =0$. 
The discrete version of this symmetry is to demand an exact $Z_2$ symmetry such that the potential is symmetric under $\Phi_1 \to \Phi_1$, $\Phi_2 \to - \Phi_2$  which leads to  $m_{12}^2=0$ and $\lambda_{6-7} =0$. This symmetry is softly broken if $m_{12}^2\neq0$, which leads to flavour changing neutral currents beyond tree-level. 
Finally in the case of a supersymmetric theory the potential takes a special form which at tree-level leads to 
\[
\begin{aligned}
\lambda_1=\lambda_2=\frac{g^2+g'^2}{4},\quad \lambda_3=\frac{g^2-g'^2}{4},\quad
\lambda_4=-\frac{g^2}{2}, \quad
\lambda_5=\lambda_6=\lambda_7=0,\quad m_{12}^2=m_A^2\cos\beta\sin\beta.
\end{aligned}
\]
For convenience the program contains two special methods for setting up: a tree-level MSSM model with parameters $m_A$ and $\tan\beta$, as well as a so called inert doublet (ID) model \cite{Barbieri:2006dq} with \{$m_h^{\rm SM}$, $m_H^{\rm ID}$, $m_A^{\rm ID}$, $m_{H^\pm}^{\rm ID}$,
 $\lambda_{2}$, $\lambda_{3}$ \} as parameters.

\section[TH constraints]{Theoretical constraints}
The 2HDMC program also contains methods for checking theoretical constraints on the parameters describing the potential, namely positiviy, perturbativity and tree-level unitarity.
    
Demanding that the potential is bounded from below one has the constraints 
\cite{ElKaffas:2006nt} :
$ \lambda_1>0$, $ \lambda_2>0$, $ \lambda_3>-\sqrt{\lambda_1\lambda_2} $.
In addition, in the case   $\lambda_6=\lambda_7=0$  one has
$\lambda_3+\lambda_4-|\lambda_5|>-\sqrt{\lambda_1\lambda_2}$
whereas if
 $\lambda_6,\lambda_7\neq0$ one gets $\lambda_3+\lambda_4-\lambda_5>-\sqrt{\lambda_1\lambda_2}$ as well as more complicated constraints which take too much room to include here, but are included in the program.

The constraints from perturbativity are obtained by considering the  cross-sections 
for $2\to2$ Higgs scattering processes,  which can be written as a perturbative series in the square of the corresponding quartic couplings $ \lambda_{H_iH_jH_kH_l}^2/(16\pi^2)$. In order for this series to make sense the couplings cannot be too large and the program contains a method for checking this with the default limit being  $\lambda_{H_iH_jH_kH_l}< 4\pi$.

A similar constraint on the parameters of the Higgs potential is obtained by 
    requiring tree-level unitarity \cite{Ginzburg:2005dt}
 for $HH$ and $HV_L$ scattering (with $V_L$ denoting the longitudinal parts of $W$ or $Z$), which corresponds to putting limits on the
     eigenvalues $\Lambda_{(Y,\Sigma)}$ of the following scattering matrices  
\begin{equation}
\begin{aligned}
16\pi S_{(2,1)}&=\left(
\begin{array}{ccc}
\lambda_1 & \lambda_5 & \sqrt{2}\lambda_6 \\
\lambda_5 & \lambda_2 & \sqrt{2}\lambda_7 \\
\sqrt{2}\lambda_6 & \sqrt{2}\lambda_7 & \lambda_3+\lambda_4
\end{array}
\right)\\
16\pi S_{(2,0)}&=\lambda_3-\lambda_4\\
16\pi S_{(0,1)}&=\left(
\begin{array}{cccc}
\lambda_1 & \lambda_4 & \lambda_6 & \lambda_6\\
\lambda_4 & \lambda_2 & \lambda_7 & \lambda_7\\
\lambda_6 & \lambda_7 & \lambda_3 & \lambda_5\\
\lambda_6 & \lambda_7 & \lambda_5 & \lambda_3
\end{array}
\right) \\
16\pi S_{(0,0)}&=\left(
\begin{array}{cccc}
3\lambda_1 & 2\lambda_3+\lambda_4 & 3\lambda_6 & 3\lambda_6\\
2\lambda_3+\lambda_4 & 3\lambda_2 & 3\lambda_7 & 3\lambda_7\\
3\lambda_6 & 3\lambda_7 & \lambda_3+2\lambda_4 & 3\lambda_5\\
3\lambda_6 & 3\lambda_7 & 3\lambda_5 & \lambda_3+2\lambda_4
\end{array}
\right) \nonumber
\end{aligned}
\end{equation}
The default value for the limit is $ | \Lambda_{(Y,\Sigma)}|<16 \pi$.

\section[Yukawa sector]{Yukawa sector}

The program contains a general Yukawa sector parameterized in terms of the matrices $\rho^L$, $\rho^D$, and  $\rho^U$ in flavour space for leptons, down-type, and up-type quarks respectively:
 \begin{equation}
  \begin{aligned}
    -\mathcal{L}_{\rm{Y}}=&\overline{D}\frac{\kd\sba+\rd\cba}{\sqrt{2}} Dh
    +\overline{D}\frac{\kd\cba-\rd\sba}{\sqrt{2}}DH+\mathrm{i}\overline{D}\gamma_5\frac{\rd}{\sqrt{2}} DA \\
    +&\overline{U}\frac{\ku\sba+\ru\cba}{\sqrt{2}}Uh
    +\overline{U}\frac{\ku\cba-\ru\sba}{\sqrt{2}}UH- \mathrm{i}\overline{U}\gamma_5\frac{\ru}{\sqrt{2}} UA \\
    +&\overline{L}\frac{\kl\sba+\rl\cba}{\sqrt{2}}Lh
    +\overline{L}\frac{\kl\cba-\rl\sba}{\sqrt{2}}LH+ \mathrm{i}\overline{L}\gamma_5\frac{\rl}{\sqrt{2}} LA \\
    +&\Bigl[\overline{U}\bigl\{V_{\rm{CKM}}\rd P_R-\ru V_{\rm{CKM}} P_L\bigr\}DH^+ + \overline{\nu}\rl P_RL H^+ + \rm{h.c.}\Bigr] \nonumber
\end{aligned}
\end{equation}
where $P_{R/L}=(1\pm\gamma_5)/2$ and $\kappa^F\equiv\sqrt{2}{M^F}/{v}$ are diagonal matrices determined by the fermion masses. The only constraint on $\rho^F$ is that they are symmetric.
 
Models with non-diagonal $\rho^F$ will lead to non-minimal flavour violating charged currents as wells as flavour changing neutral currents. This can be avoided  \cite{Glashow:1976nt}
 by imposing a $Z_2$ symmetry on $\Phi_1$, $\Phi_2$ and $U_R$, $D_R$, $L_R$ such that each fermion type only couples to one Higgs doublet. In turn this leads to the restriction
 $\rho^F=\kappa^F\cot\beta$ or $\rho^F=-\kappa^F\tan\beta$, which means that there are
 four different types of  $Z_2$ symmetric 2HDMs. For simplicity the program also contains methods for specifying the Yukawa sector by simply choosing the type. 

 \section[Widths]{Partial decay widths}
With both the Higgs sector and the Yukawa sector being specified the program can then be used to calculate partial decay widths and branching ratios for range of processes including:
        \begin{itemize}
       \item 
       $H\to ff^\prime$ with optional (N)LO QCD corrections 
       \item 
       $H\to gg$ with optional LO QCD corrections 
       \item 
       $H\to HH$ 
       \item 
       $H\to HV^*$ including off-shell vector bosons
       \item 
       $H\to VV^*$  including off-shell vector bosons
       \item 
       $H\to \gamma \gamma$
       \item 
       $t \to H^+ b$
       \end{itemize}
where $H=\{h,H,A,H^\pm\}$ denotes any applicable Higgs boson and $V=\{Z,W\}$.

\section[EX constraints]{Experimental constraints}

Before using the program to make predictions in a specific model one may also want to take into account existing experimental constraints. This is not a simple task and there are dedicated programs for this. Therefore, the program has been limited to having methods for calculating the 2HDM contribution to the oblique parameters \cite{Grimus:2008nb}:  $S$, $T$, $U$, $V$, $W$, $X$  compared to the SM with Higgs mass $m_h^{\rm ref}$,   as well as the contribution to the  muon anomalous magnetic moment. 
In order to take into account collider and flavour limits the program contains interfaces to  HiggsBounds (version 2)~\cite{Bechtle:2008jh}  and SuperIso~\cite{Mahmoudi:2008tp} respectively.

\section[Usage]{Usage/parameter-setting/input-output}

The 2HDMC program has been written in a modular fashion with an object oriented structure using C++. It can either be used in library mode where the different methods are called by the user program or as ``ready to compile" command line type programs which are distributed together with the main code.

The code, manual, and full class documentation can be downloaded from \url{http://www.isv.uu.se/thep/MC/2HDMC}. In order to compile the program one needs a gcc compiler (3.4 and 4 tested) and the GNU Scientific Library (GSL) installed. There is a template makefile which should be adapted to the local conditions and there is a test program {\tt Demo} with which the successful installation can be tested.

The code consist of four main classes
\begin{itemize}
\item
{\bf SM}:	to give the Standard Model input parameters with running $\alpha_s$ and quark masses
\item
{\bf THDM}:	to specify a general two-Higgs doublet model in terms of the Higgs and Yukawa sectors
\item
{\bf Constraints}: to calculate theoretical and experimental constraints on the 2HDM	 
\item
{\bf DecayTable}: to calculate the decay modes of 2HDM Higgs bosons and the top quark
\end{itemize}
For a more complete description we refer to the class documentation which can be found on the 2HDMC homepage.

In order to simplify the communication with other programs there is also a LesHouches style input/output format and one can also export a given 2HDM as a user model for
 MadGraph/MadEvent~\cite{Alwall:2007st}
 to calculate cross-sections etc.

\section{Summary and conclusions}

The 2HDMC program is a versatile and flexible program for exploring the physics of general CP-conserving 2HDMs: all the way from defining the model and checking it against theoretical and experimental constraints, to exploring its phenomenology such as calculating decay widths and scattering cross-sections. Some examples of its use can be found in \cite{Mahmoudi:2009zx,Bernreuther:2010uw,Moretti:2010kc,Enberg:2011qh}.

\end{document}